\documentclass[12pt]{article}
\usepackage[english]{babel}
\usepackage{blindtext}
\usepackage[T1]{fontenc}
\usepackage[utf8]{inputenc}
\usepackage{float}
\usepackage{textcomp}
\usepackage{amsmath}
\usepackage{amssymb}
\usepackage{graphicx}
\usepackage{siunitx}
\usepackage[dvipsnames]{xcolor}
\usepackage{ulem}
\usepackage{natbib}
\setcitestyle{super,square,comma}

\usepackage{lipsum, babel}

\usepackage[symbol]{footmisc}
\renewcommand{\thefootnote}{\fnsymbol{footnote}}
\newenvironment{sciabstract}{%
\begin{quote} }
{\end{quote}}
\newcommand\blfootnote[1]{%
  \begingroup
  \renewcommand\thefootnote{}\footnote{#1}%
  \addtocounter{footnote}{-1}%
  \endgroup
}

\title{Enhanced polariton interactions in suspended WS$_2$ monolayer microcavity\\ 
}
\author{\textit{Laura Polimeno}$^{1 \dagger \ast}$, \textit{Francesco Todisco}$^{1 \dagger \ast}$,
\textit{Rosanna Mastria}$^{1}$,\\ \textit{Milena De Giorgi}$^{1}$, \textit{Antonio Fieramosca}$^{1}$, \textit{Marco Pugliese}$^{1}$, \\ \textit{Dario Ballarini}$^{1}$, \textit{Anna Grudinina}$^{2,3}$, \textit{Nina Voronova}$^{2,3}$,\\ \textit{Daniele Sanvitto}$^{1}$
\\\\
\normalsize{$^{1}$CNR Nanotec, Institute of Nanotechnology,}
\normalsize{via Monteroni, 73100, Lecce, Italy.}
\\
\normalsize{$^{2}$National Research Nuclear University MEPhI, (Moscow Engineering Physics Institute),}\\
\normalsize{Kashirskoe shosse 31, 115409 Moscow, Russia.}\\
\normalsize{$^{3}$Russian Quantum Center, Skolkovo IC, Bolshoy boulevard 30 bld. 1, 121205 Moscow, Russia.}
}
\date{}

\begin{document}

\baselineskip24pt
\maketitle

\blfootnote{$^{\dagger}$ These authors contributed equally: L.~Polimeno, F.~Todisco\\
$^{\ast}$ e-mail: laura.polimeno@nanotec.cnr.it, francesco.todisco@nanotec.cnr.it}

\section*{Abstract}

\begin{sciabstract}
Transition-metal dichalcogenides monolayers exhibit strong exciton resonances that enable intense light-matter interactions at room temperature (RT). However, the sensitivity of these materials to the surrounding environment and their intense interactions with the sustaining substrate result in the enhancement of excitonic losses through scattering, dissociation and defects formation, hindering their full potential for the excitation of optical nonlinearities in exciton-polariton platforms. From this point of view, the use of suspended monolayers holds the potential to completely eliminate substrate-induced losses, offering unique advantages for the investigation and exploitation of intrinsic electronic, mechanical, and optical properties of 2D materials-based polaritonic systems, without any influence of proximity effects of all sort.\\
In this work, we report a novel fabrication approach enabling the realization of a planar $\lambda/2$ microcavity filled with a suspended WS2 monolayer in its centre. In such a system, we experimentally demonstrate a 2-fold enhancement of the strong coupling at RT, due to reduced overall losses as compared to similar systems based on dielectric-filled microcavities. Moreover, as a result of minimized losses, spin-dependent polaritonic interactions in our platform are significantly amplified, leading to achievement of a record exciton interaction constant approaching the theoretically predicted value at RT, without making use of theoretical hypothesis on the effective polariton densities.\\
This approach holds promises for pushing 2D materials-based polaritonic systems to their intrinsic limits, paving the way for the realization of novel polaritonic devices with superior performance.

\end{sciabstract}

\section*{Introduction}
Exciton polaritons are hybrid quasi-particles arising from the strong coupling between photons and excitons in semiconductors\cite{kavokin2017microcavities}, and represent a promising frontier for the development of next-generation optoelectronic and quantum technologies, working at room temperature (RT) \cite{sanvitto2016road,ghosh2022microcavity}.
Among all the unique properties of these hybrid light-matter states, their strong nonlinear character represents one of the most interesting features, since it opens the way to the observation of many fascinating effects, including superfluidity \cite{lerario2017room,amo2009superfluidity}, Bose-Einstein condensation \cite{kasprzak2006bose} and long-range energy transport \cite{rozenman2018long}. \\
Recently, atomically thin transition-metal dichalcogenides (TMDs) have captured considerable interest from different solid state research communities, including photonics and polaritonics, due to their distinctive optical and structural characteristics \cite{Manzeli20172DDichalcogenides,Jariwala2014EmergingDichalcogenides}. In fact, TMD monolayers exhibit strong quantum confinement and reduced dielectric screening, leading to large excitonic binding energies (0.6 -- 1 eV) that make excitonic resonances robust even at RT \cite{Ramasubramaniam2012LargeDichalcogenides,Zhu2015ExcitonWS2,Cheiwchanchamnangij2012Quasiparticle2}. The atomic thickness of these materials and their defined in-plane oriented dipole, moreover, makes them an ideal active material for probing strong light-matter interactions in different platforms, ranging from planar microcavities \cite{anton2021bosonic, zhao2022nonlinear, zhao2024room} to waveguides \cite{wu2023strong}, metasurfaces \cite{weber2023intrinsic}, Bloch surface waves \cite{maggiolini2023strongly} and plasmonic and dielectric nanostructures\cite{geisler2019,sortino_enhanced_2019}.\\
The principal constraint in employing TMD monolayers as active media in exciton polariton platform is their high interaction with substrates, interfaces, and overall surrounding environment, all contributing to a reduction of the oscillator strength and quenching of the photo-emission\cite{kim2018photoluminescence, pan2022exciton,PRBExcitonDissociation}. In fact, upon depositing a TMD monolayer onto a substrate, interface effects induce the formation of defect-mediated localized states, doping effects or uncontrollable strain \cite{liu2020substrate,van2023impact}, introducing non-radiative losses channels and localised exciton traps.\\
Until now, different fabrication techniques have been employed to minimize the effects of interactions with the substrate, such as encapsulation in multilayer hexagonal boron nitride (hBN)\cite{han2019effects, lee2023suppression}, the use of self-assembled monolayers on various substrates \cite{park2024mitigating} and passivation treatments \cite{tanoh2019enhancing}. All these processes, however, rely on the inclusion of additional materials to induce surface modification of the dielectric environment surrounding TMD monolayers, thus suffering from spatial inhomogeneities and introducing additional spacing around the monolayer, that can locally screen near-field interactions and add proximity effects. In this sense, an alternative and promising approach involves completely suspending the monolayer as a free-standing membrane, thereby eliminating substrate-induced effects \cite{huang2022efficient,hernandez2022strain, sun2022enhanced} and providing an optimal platform for investigating excitonic properties in the strong coupling regime in microcavities.\\
In this study, we report a reliable fabrication process for embedding a suspended tungsten disulfide (WS$_2$) monolayer in the middle of a planar microcavity, allowing us to explore the ultimate limits of TMD-based microcavity polaritons, free from any near-field interactions with the bottom and top mirrors. In such a configuration, we theoretically simulate and experimentally demonstrate strong light-matter interactions of the suspended monolayer, showing higher oscillator strength as compared to dielectric-filled microcavities and enhanced optical nonlinearities and polariton-exciton interactions. By completely suppressing substrate-induced effects, the excitonic losses are minimized and a record Rabi splitting of 56 meV is obtained. Moreover, using both conservative and direct experimental methods to determine the polariton density, the nonlinear interactions in the suspended monolayer-based cavity are significantly stronger than those reported for WS$_2$ monolayers at RT, demonstrating not only the maximization of the intrinsic oscillator strength of these materials but also the preservation of spin-dependent interactions.\\
Our results represent an initial step towards the realization and fabrication of all-optical polariton platforms that fully exploit the optical properties of TMD monolayers, overcoming the intrinsic limitations encountered until now in these systems due to proximity effects. Additionally, we emphasize that the suspended region can be easily prepared with specific geometrical patterns, paving the way for the straightforward realization of various types of suspended structures in structured artificial potentials.

\section*{Results and Discussion}
\begin{figure}[h!]
\centering \includegraphics[scale=0.35]{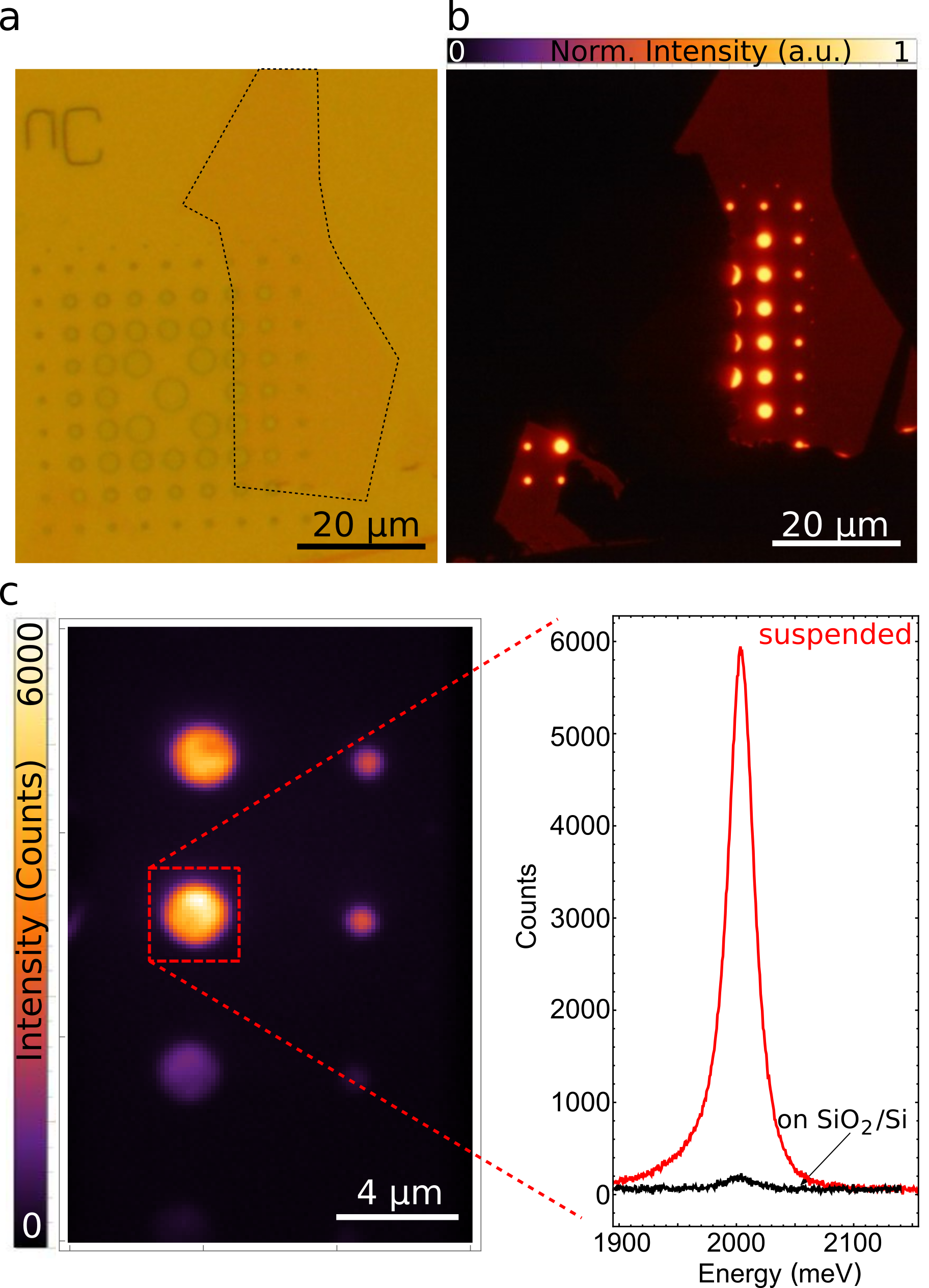}
\caption {Real space bright-field reflection (a) and photoluminescence (b) image of a suspended WS$_2$ monolayer at RT. The black dashed line in (a) indicated the contour of the monolayer flake transferred on the holes array. The dimension of the holes ranges from 1 $\mu$m to 5 $\mu$m. c) Zoom of the monolayer photoluminescence on a 2 $\mu$m hole. The monolayer is off-resonantly pumped with a CW laser at $\sim 2541$~meV, with a 100~$\mu$m spot.  d) Comparison of the monolayer photoluminescence spectra recorded from suspended (red) and SiO$_2$/Si substrate (black) regions.}
\label{fig1} 
\end{figure}
To demonstrate the improvement and enhanced performance of our suspended system, we selected WS$_2$ from the range of TMD monolayers because in the encapsulated configuration it has been shown to exhibit weaker exciton interaction (g$_{exc}\sim$ 0.2 $\mu$eV $\mu$m$^2$ at RT)\cite{zhao2024room, barachati2018interacting, zhao2022nonlinear, luo2023manipulating}.\\
Firstly, to investigate the optical properties of suspended WS$_2$ monolayer membranes, we etched 300 nm deep holes onto a commercial Si/SiO$_2$ substrate, with diameters ranging from 1 $\mu$m to 5 $\mu$m (see Section 1 of the Supporting Information for further details). A WS$_2$ monolayer is then mechanically exfoliated and transferred onto the patterned area via the dry transfer method \cite{Castellanos-Gomez2014}, followed by a vacuum hard baking on a hot plate to improve adhesion (3 hours at 200$^{\circ}$ C).\\
Fig.~\ref{fig1}a shows the bright-field reflection image of the monolayer (black dashed line) deposited on the holes array. The corresponding photoluminescence (PL) image in true colors, recorded with a colour CCD camera on an optical microscope equipped with a blue LED light source and a coloured filter, is shown in Fig.~\ref{fig1}b, perfectly following the holes pattern and demonstrating the enhancement of the PL in the suspended area.\\
To quantitatively evaluate the PL enhancement, the sample is off-resonantly excited with a continuous wave laser (energy peak at 2541 meV) with a spot size of $\sim$ 100 $\mu$m, exciting both the free-standing and substrate-attached regions, as shown in Fig.~\ref{fig1}c. Under the same pumping conditions (excitation density $\sim$ 10 nW/$\mu$m$^2$), the PL intensity from the suspended WS$_2$ is 10 times stronger than that from the attached one. Additionally, the PL lineshape of the suspended WS$_2$ is narrower due to a decrease in the trion contribution to the PL, attributed to a reduction in unintentional doping ad strain (Fig. S1 of the Supporting Information). This difference in photoemission efficiency clearly demonstrates that doping and defects-related effects are mostly suppressed in the suspended monolayer, which is completely isolated from the environment \cite{huang2022efficient, hernandez2022strain}. The excitonic resonance in the suspended region is free from substrate-induced effects, with minimized losses. As a consequence, free-standing monolayer membranes appear to be ideal active platforms for improving strong light-matter interactions in planar microcavities.
\begin{figure}[h!]
\centering \includegraphics[scale=0.75]{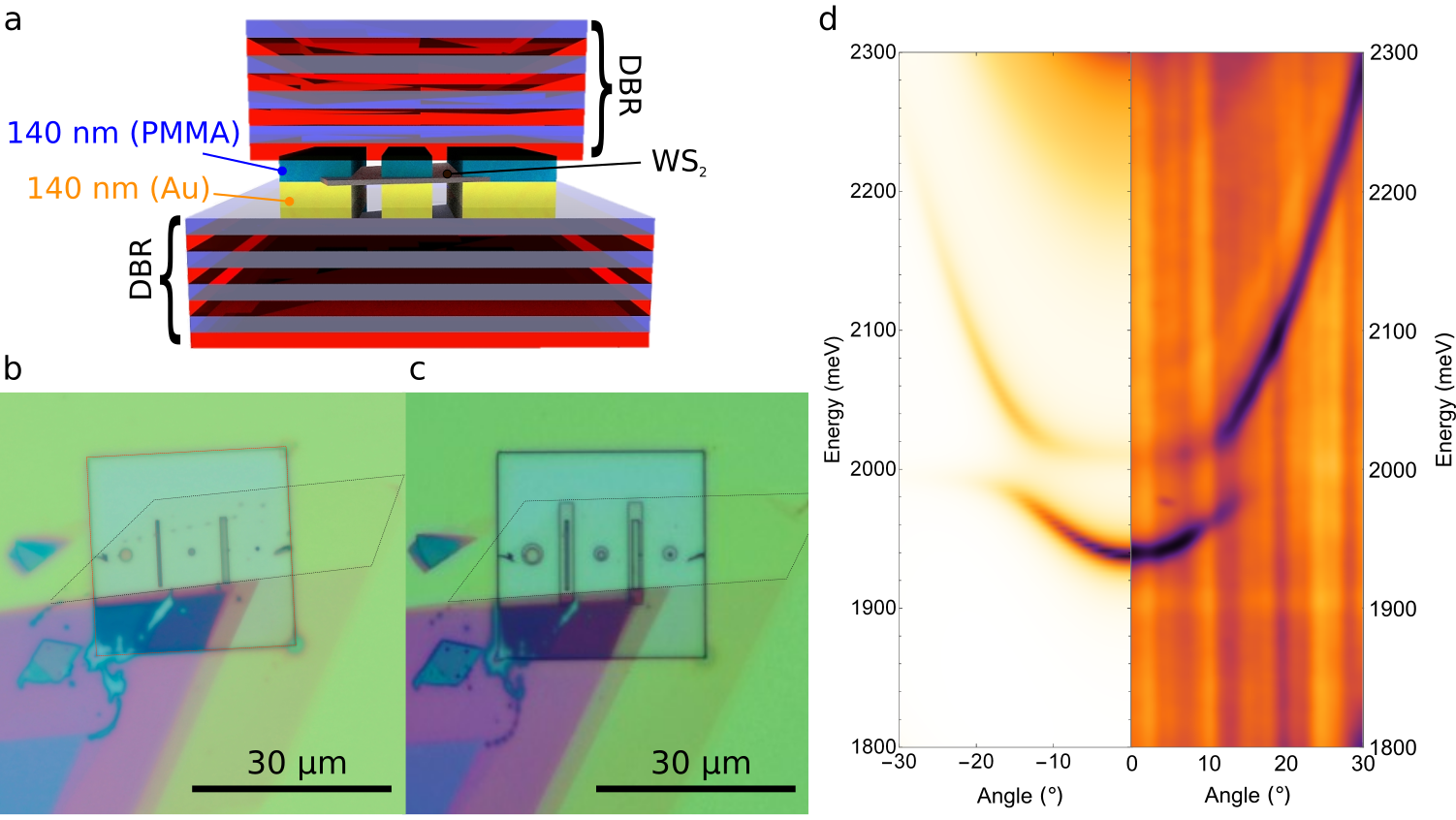}
\caption{a) Sketch of the planar microcavity. The 140 nm-thick gold island is fabricated on top of a DBR made of 4 pairs of SiO$_2$/TiO$_2$ layers. The WS$_2$ monolayer is suspended inside the planar microcavity at the position where the confined electromagnetic field is maximum. b) Real space image of the sample after mechanical transfer. The monolayer flake is transferred onto two empty stripes of 1x10 $\mu$m and 0.5x10 $\mu$m, etched into the gold island. c) Real space image of the sample after writing the top PMMA using EBL. The top stripes are larger than the bottom ones to facilitate the alignment of the suspended region. d) Comparison between the theoretical simulation (left-hand side) and the experimental reflectivity (right-hand side) in Fourier space of the 1x10 $\mu$m stripe. The white light (spot dimension $\sim$ 1 $\mu$m) is focused on the suspended region.}
\label{fig2} 
\end{figure}

The design of our planar microcavity is illustrated in Fig.~\ref{fig2}a (all fabrication steps are described in Section 2 of the Supporting Information), where the top and bottom mirrors consist of two Distributed Bragg Reflectors (DBRs), each made of 4 pairs of SiO$_2$/TiO$_2$ layers. In such a system, the most crucial parameter for maximizing light-matter interactions is the position of the free-standing membrane, that needs to be placed at the maximum of the confined electromagnetic field. In particular, by using high refractive index terminated DBRs, we found the optimum air-filled exciton-resonant microcavity total thickness to be 280 nm, corresponding to a $\lambda$/2 cavity.\\
In order to manifacture such a platform, a 30 $\mu$m x 30 $\mu$m gold pad with two rectangular holes of 1x10 $\mu$m and 0.5x10 $\mu$m is fabricated on the surface of a DBR by electron beam lithography (EBL) followed by 140 nm metal evaporation (3 nm chromium followed by 137 nm gold) and lift-off (see Methods for details). A WS$_2$ monolayer is then mechanically transferred on the structured pad (as highlighted by a black dashed line in Fig.~\ref{fig2}b), followed by vacuum hard baking to improve adhesion (Fig. S3 of the Supporting Information). Then, a 140-nm thick Polymethyl-methacrylate (PMMA) top layer is spin coated on the sample and two stripes are defined by electron beam lithography on top of the free-standing monolayer regions. In order to maximize the suspended area and minimize alignment errors, the top stripes are intentionally defined larger than the bottom ones, as shown in Fig.~\ref{fig2}c. The microcavity is finally closed by mechanical transfer of the top DBR onto the structured area, as already reported in previous works \cite{anton2021bosonic,paik2023high,rupprecht2021micro, zhao2022nonlinear}.\\
The theoretical simulation of the Fourier space reflectivity dispersion of the designed microcavity is shown in left panel of Fig.~\ref{fig2}d. The energy of the optical mode is tuned to the WS$_2$ excitonic resonance ($\sim 2000$~meV), but strong coupling is achieved only in the suspended region, where the maximum of the electromagnetic field coincides with the spatial position of the monolayer (see Section 3 of the Supporting Information). 
The right half of Fig.~\ref{fig2}d shows the measured angular dispersion of the reflectivity for the 1x10 $\mu$m stripe. The long dimension of the stripe is aligned with the entrance of the slit, and a focused white light (spot size reduced to 1 $\mu$m) is used to select only the signal coming from the suspended region. The angular dispersion clearly shows the bending of the optical mode at the excitonic resonance (Exc $\sim$ 2000 meV) at angle of $\theta$ $\sim$ 15$^\circ$, in good agreement with the simulation. As theoretically predicted, the planar microcavity is perfectly tuned with the WS$_2$ exciton in the entire suspended region, while only the optical mode is measurable in the substrate-attached area (Fig. S4 in Section 3 of the Supporting Information).

\begin{figure}[h!]
\centering \includegraphics[scale=0.6]{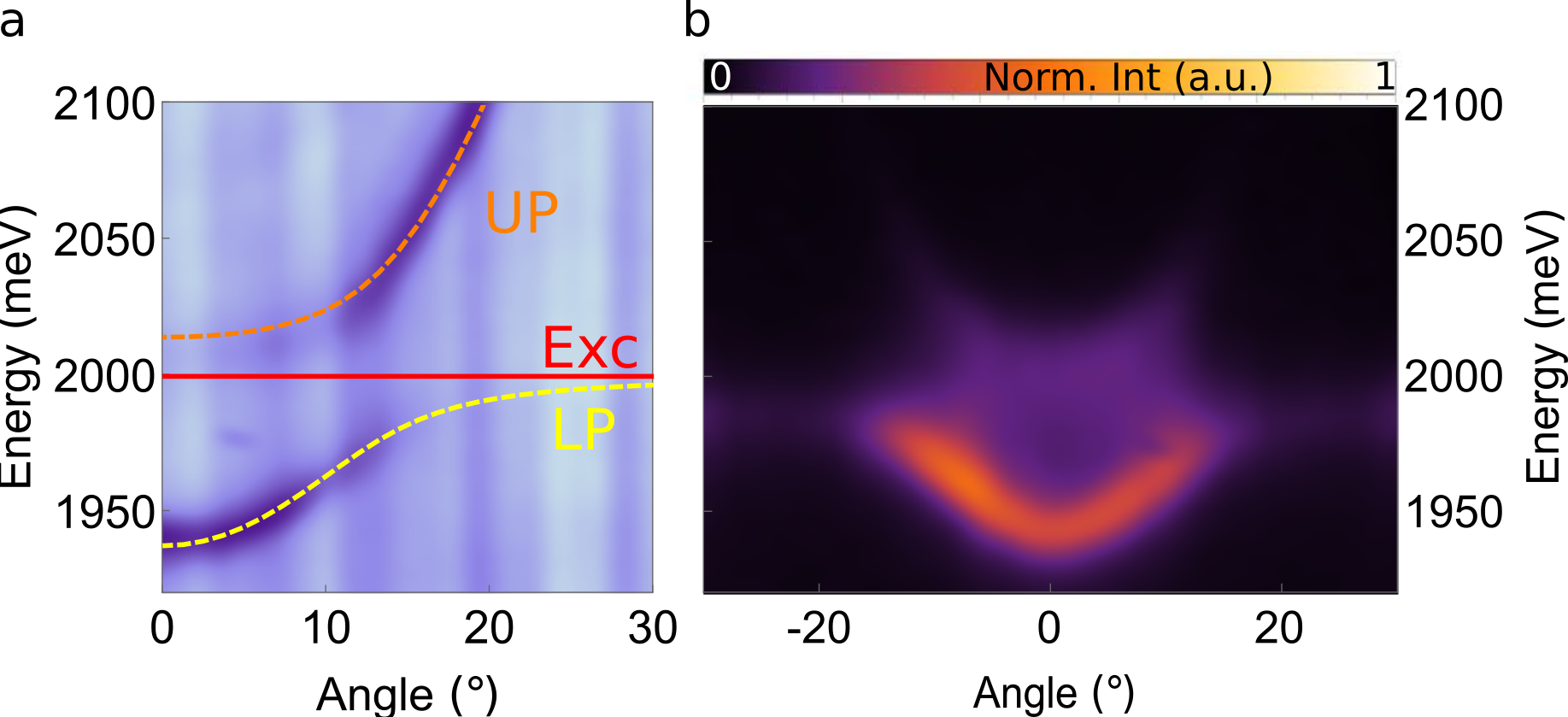}
\caption{a) Fitting of the experimental reflectivity with a two coupled oscillators model. The strong coupling between the excitonic resonance (Exc) of the suspended monolayer  (cointinuous red line) and the optical mode confined in the cavity leads to the formation of the upper polariton (orange dashed line) and lower polariton (yellow dashed line) branches, with a Rabi splitting of 56 meV. b) Measured angular dispersion photoluminescence of WS$_{2}$ suspended monolayer in the planar microcavity. }
\label{fig3} 
\end{figure}

The anticrossing of the optical mode with the monolayer exciton is fitted using a two-coupled oscillators model (see Methods for futher information), as shown Fig.~\ref{fig3}a. The dashed yellow and orange lines represent the fitting of the lower and upper polariton branches, respectively, extracting a Rabi splitting of $\Omega_R$ $\approx 56$~meV. The $\Omega_R$ found for the suspended monolayer in the strong coupling regime is twice that of a typical planar microcavity where the monolayer is embedded within two dielectric-filled DBRs \cite{zhao2021ultralow,zhao2022exciton,lackner2021tunable}, demonstrating a significant reduction in excitonic losses in our system and a strong enhancement of the oscillator strength.

Off-resonantly pumping the system with a CW laser (peak energy at $\sim$ 2541 meV), we observe photoemission from the lower polariton branch that appears in the angle-resolved dispersion (Fig.~\ref{fig3}b), demonstrating the enhancement of the PL efficiency by reducing the substrate interaction. In our suspended system in the strong coupling regime, interactions between the monolayer and other surfaces are completely eliminated, enhancing its interaction with the optical mode confined in the microcavity. The air layers at the bottom and top effectively isolate the exciton from uncontrolled doping, strain, and the formation of defects, which are commonly present in typical SiO$_2$- or PMMA-filled microcavities.\\
\begin{figure}[h!]
\centering \includegraphics[scale=0.35]{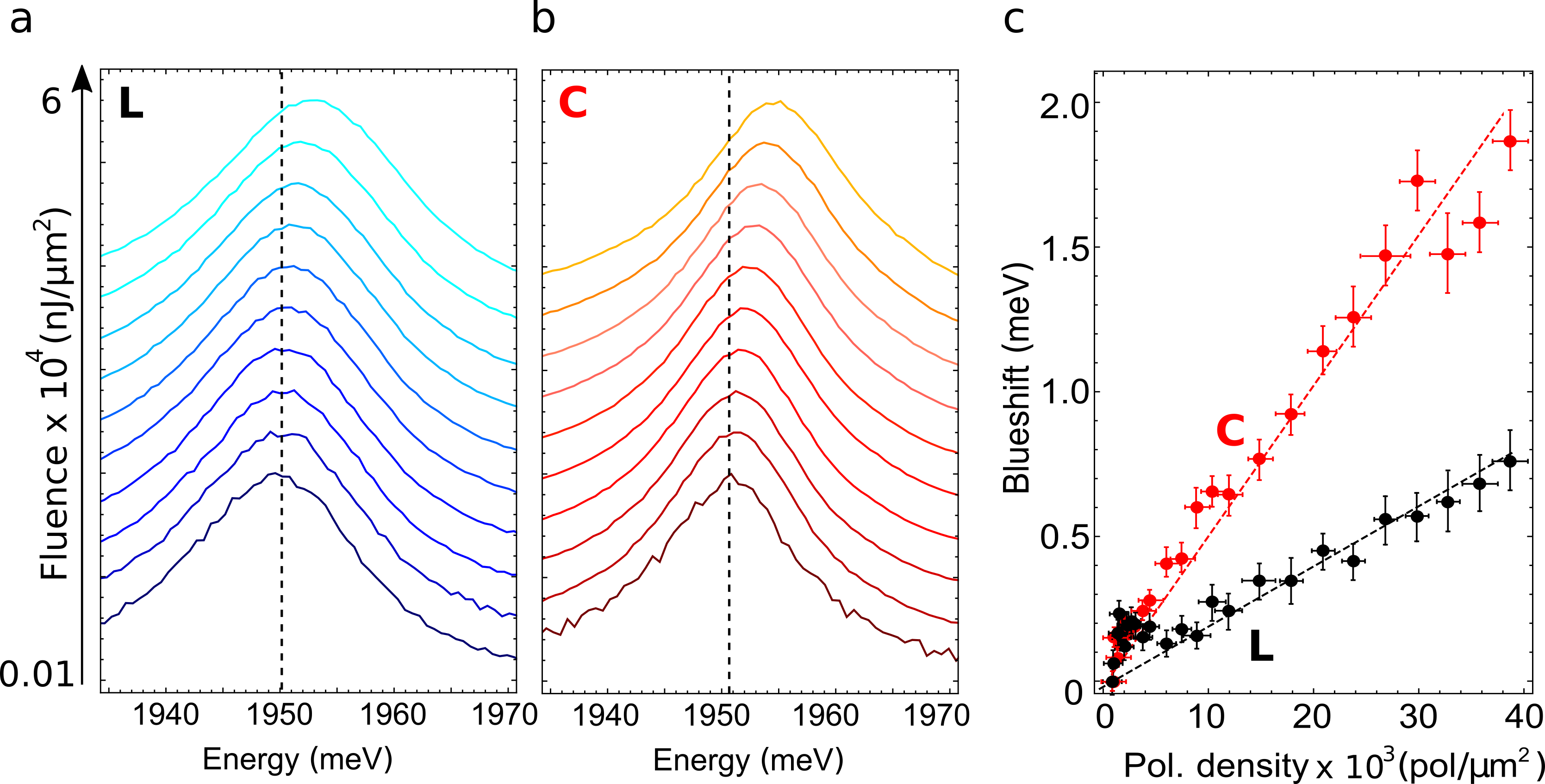}
\caption{a), b) Transmitted spectra of the pulsed laser on polariton mode at the angle $\sim 0^\circ$, corresponding to different resonant pumping fluences for linearly- (a) and circularly- (b) polarized excitation laser. c) Energy blueshift of the polariton modes at $\theta$ $\sim 0^\circ$ as a function of the polariton density inside the cavity, in the case of linearly- (L, black dots) and circularly- (C, red dots) polarized 
excitation. The dashed lines are linear fits to the experimental data, with the slope yielding the polariton interaction constant, $g_{\text{pol, lin}} = 0.017~\mu$eV$\cdot\mu$m$^2$ for the linear case and $g_{\text{pol, circ}} = 0.046~\mu$eV$\cdot\mu$m$^2$ for the circular one.}
\label{fig4} 
\end{figure}
To further investigate the effects of the reduction of substrate scattering and losses on the polaritons optical properties, we measured the polarization-dependent polariton-polariton interactions in our suspended platform. The underlying mechanism governing nonlinear interactions in TMD monolayer-based exciton-polaritons is highly debated, with different results and interpretations reported in literature to date \cite{stepanov2021exciton, choo2024polaronic, tan2023bose}. However, it can be expect that reduced substrate scattering can improve the polarization coherence of polaritons, leading to stronger polarization-dependent nonlinearities when resonantly pumping with circularly polarized light. In fact, using a linearly polarized excitation---which is a coherent mix of two oppositely polarized circular polarizations---only half of the excited polaritons will have the same spin, which effectively reduces the interaction energy by approximately a factor of two.\\
To investigate this effect, the lower polariton branch on the suspended region was resonantly pumped at $\theta\sim0^\circ$ with a resonant fs-pulsed laser in transmission configuration, at different incident fluences, with the incident laser linear and circular polarization. The corresponding fluence-dependent transmission spectra are shown in Fig.~\ref{fig4}a and ~\ref{fig4}b, respectively. In both cases, as the fluence increases, the lower polariton branch continuously shifts towards higher energies (blueshifts) due to the the combined effect of the increasing  polariton interactions and the so-called exciton saturation (often referred to phase-space filling~\cite{tassone,glazov}). \\
We note that at higher fluences, the increase of excitons and photons number leads to a higher density of polaritons in the system. The measured blueshift thus allows the estimation of the lower-polariton nonlinearity, if an accurate knowledge of the polariton density can be acquired. To experimentally evaluate this density in our system, we directly measured the transmitted laser fluence $F$ passing through the microcavity defined by an empty (without suspended TMD monolayer) stripe with a rectangular shape of 1 $\mu$m x 5 $\mu$m. Consequently, the upper bound for the polariton density can be defined as $F/E_{ph}$, where the energy of the resonant pumping laser is $E_{ph}\sim 1950$~meV. It is important to note that this model assumes that all photons entering the system are converted into polaritons, providing 
an upper limit for the maximum polariton density as function of the fluence. \\ 
Defining the blueshift 
of the polariton dispersion as  $\Delta E_{pol} = g_{pol}\cdot n_{pol}$,
where $g_{pol}$ is the effective strength of the polariton nonlinearity, by linear fitting of the data reported in Fig.~\ref{fig4}a,b as a function of the experimental polariton density (Fig.~\ref{fig4}c) we estimate 
at RT $g_{pol,lin} \approx 0.017~\mu$eV $\mu$m$^2$ and $g_{pol,circ} \approx 0.046~\mu$eV~$\mu$m$^2$ for the linearly and circularly polarized excitation, respectively. 
The energy blueshift achieved with circularly polarized excitation is thus substantially higher than that observed when exciting with a linearly polarized laser. This implies that the energy blueshift is substantially larger when all polaritons have the same spin, as expected for a more strongly interacting polariton system \cite{ciuti2000theory, vladimirova2010polariton, vladimirova2009polarization}.\\
This result aligns with the typical behavior of polariton-polariton interactions in standard semiconductors at cryogenic temperatures, characterized by a strong repulsive interaction for same-spin polaritons and a weaker interaction for opposite-spin polaritons \cite{vladimirova2009polarization}.\\
Note that spin-dependent interactions are completely suppressed in a typical dielectric-filled WS$_2$ microcavity, as shown in Fig. S5 and in Section 4 of the Supporting Information. It has been anticipated that in this case the main mechanism could be phonon-mediated interactions induced by the presence of the substrate, which are spin-independent. 
As we show here, spin-anisotropic interactions are restored when a suspended monolayer is considered, possibly because phonon scattering is reduced in this configuration.
This comparison demonstrates that eliminating uncontrollable substrate interactions is essential for preserving spin-dependent interactions. \\
Generally, in the mean field approximation the lower-polariton blueshift~\cite{grudinina2024} $\Delta E_{pol}=g_{pol}\cdot n_{pol}$ contains two contributions since $g_{pol} = g_{exc}|X|^4 + 2g_{sat}|C||X|^3$, where $g_{exc}$ and $g_{sat}$ are the exciton-exciton interaction and exciton saturation constants (see Eq.~(S9)--(S10) in Section 5 of the Supporting Information), $X$ and $C$ are the exciton and photon Hopfield coefficients, respectively, here defined for zero wavevector. At the same time, our theory provides estimates for $g_{sat}$ to be an order of magnitude smaller than $g_{exc}$ in all range of temperatures (see Section 5 of the Supporting Information), therefore we can conclude that the nonlinearity is mainly governed by interactions. 
With the detuning of the pumped mode $\delta \approx -50$~meV and the observed Rabi splitting $\Omega \approx 56$~meV, the excitonic fraction of the lower polariton branch at $\theta = 0^\circ$ is $|X|^2=0.167$. Extracting the excitonic interaction constant at RT we finally get $g_{exc,lin}\approx 0.61~\mu$eV~$\mu$m$^2$ and $g_{exc,circ}\approx 1.65~\mu$eV~$\mu$m$^2$, for linearly and circularly polarized excitation, respectively. The exciton interaction constant $g_{exc}$ extracted from this analysis neglects any material losses, thereby representing a lower bound for the interaction strength. Despite this, the $g_{exc, circ}$ value extracted for the suspended WS$_2$ at RT is at least six times higher than the $g_{exc}$ experimentally extracted for both a standard WS$_2$ monolayer-based microcavity (see Section 4 of the Supporting Information) \cite{zhao2022nonlinear,luo2023manipulating} and also for other polariton platforms based on TMDs \cite{maggiolini2023strongly, barachati2018interacting}.

Moreover, interactions measured in our case are additionally enhanced by RT. The theoretical value $g_{exc}^{\rm RT}=2.0804~\mu$eV~$\mu$m$^2$ (see Section 5 of Supporting Information for calculation details) for the same-spin species appears to be 1.6 times larger than the estimate for $T=0$, and is in excellent agreement with the lower-bound value $1.65~\mu$eV~$\mu$m$^2$ obtained for our suspended WS$_2$ microcavity when excited with circularly-polarized laser. We note that our experimental estimation likely overstates the polariton density by neglecting, among other things, also the exciton-exciton annihilation, therefore it corresponds to a conservative evaluation of the interaction strengths.

We demonstrate that our suspended polariton platform at RT overcomes the limitations of typical polariton systems fabricated until now---completely suppressing excitonic losses caused by interaction with the environment,---strongly enhancing polariton interactions. 

\section*{Conclusion}
Investigating the optical properties of suspended WS$_2$ monolayers is essential for optimizing and enhancing exciton-polariton interactions in TMD-based microcavity.\\
In this work, we firstly demonstrate how to enhance the excitonic emission of WS$_2$ by suspending the monolayer as a free-standing membrane, thus suppressing all substrate proximity effects. This isolation resulted in a tenfold increase in PL intensity compared to contacted regions, while maintaining the PL shape.\\
Then,  we develop a novel fabrication approach to introduce a suspended monolayer in a planar microcavity, ensuring that the maximum of the confined electromagnetic field is precisely positioned at the suspended monolayer region.\\
The experimental and theoretical analysis confirmed strong coupling between the suspended WS$_2$ exciton and the optical mode, with a Rabi splitting of 56 meV, significantly higher than in traditional microcavities.
Furthermore, our system demonstrated a substantial reduction in excitonic losses and enhanced polariton interactions, as evidenced by the fourfold increase in the exciton interaction constant ($g_{exc}$) compared to standard WS$_2$ monolayer-based microcavities and exciton-polariton platform working at RT. Moreover, the spin-dependent interactions are preserved by eliminating the substrate interaction.\\
Overall, our suspended polariton platform overcomes the limitations of typical polariton systems by eliminating environmental interactions and enhancing the intrinsic properties of the WS$_2$ monolayer.\\
This improvement underscores the potential of suspended monolayers in strong coupling regime for advanced photonic applications, including topological photonics, photonic crystals, and valleytronics.

\section*{Materials and Methods}

\subsection*{Sample Fabrication}
The DBR is formed by four pairs of SiO$_2$/TiO$_2$ layers (with thicknesses of 106 nm/68 nm, respectively), deposited by electron-beam evaporation (Temescal Supersource), keeping the chamber at 10$^{-5}$ $\div$ 10$^{-6}$ mbar throughout the process, with the sample at room temperature and in absence of any oxigen gas flow (deposition rates: 1 Å/s for SiO$_2$, 0.5 \AA/s for TiO$_2$). resulting DBR stopband is centered at 2000 meV (620 nm).\\
The suspended TMD-based microcavity is defined by a two step litography process. The full details of this procedure is detailed in Section 2 of the Supporting Informations.
Single-layer WS$_2$ is mechanically exfoliated from bulk crystals (HQ Graphene) with Nitto SPV 224 tape and transferred onto the surface of a PDMS stamp (Gelfilm, retention level x4 from Gel-Pak®). Single-layer WS$_2$ is transferred by all-dry deterministic transfer\cite{Castellanos-Gomez2014} on the gold island by using a micromanipulator. The sample is finally annealed in vacuum at 200$^\circ$C for 3h.

\subsection*{Optical Measurements}

All measurements reported in this work are performed under ambient conditions at RT. \\
For photoluminescence measurements (both in real and Fourier space), the WS$_2$ monolayer is off-resonant excited by using a continuous-wave 488 nm diode laser. The photoluminesce is recorded in reflection configuration, using a 100x objective with NA=0.5.\\
The reflectivity measurements of the polariton dispersion are performed using an home-built microscope, equipped with a 0.5 numerical aperture (NA) objective with 100x magnification. Four lenses are used to project a magnified image of the back focal plane (BFP) onto the slit of an imaging spectrometer with a cooled charge-coupled device camera. \\
For nonlinear measurements, a tunable femtosecond laser (with pulse width $\sim$ 145 fs, repetition rate 10 kHz) is focused onto the BFP of the 0.55 NA objective at $\theta$ $\sim$ 0.
The energy of the laser resonantly pumping the interested polariton mode and the corresponding real space spot size dimensions are $\sim$ 1950 meV and $\sim$ 10 $\mu$m$^{2}$, respectively. 

\section*{Theoretical simulation and fits}
Simulations of the reflectivity were instead performed by an open source implementation of the RCWA method\cite{Liu20122233, maggiolini2023strongly}.\\
The experimental data reported in Fig. 3a of the main text are fitted with a 2x2 coupled harmonic oscillator Hamiltonian:
\begin{eqnarray}
H_{\mathbf{k}}=\begin{pmatrix}
E_{ph}(\theta) &\Omega_R/2 \\
\Omega_R/2& Exc
\label{Hameff}
\end{pmatrix} 
\end{eqnarray}
where $Exc$ is the exciton energy, $\Omega_R$ is the Rabi splitting and $E_{ph}(\theta)$ is the energy dispersion of the photonic branch.

\section*{Data availability}
The data that support the plots within this paper and other findings of this study are available from the corresponding authors upon reasonable request.

\section*{Acknowledgements}
The authors gratefully thank P. Cazzato for technical support.\\
This work was supported by the Italian Ministry of University (MUR) through the PNRR MUR project: ‘National Quantum Science and Technology Institute’ --- NQSTI (PE0000023), 
PNRR MUR project: ‘Integrated Infrastructure Initiative in Photonic and Quantum Sciences’ --- I-PHOQS (IR0000016), Quantum Optical Networks based on Exciton-polaritons -- (Q-ONE) funding from the HORIZON-EIC-2022-PATHFINDER CHALLENGES EU programme under grant agreement No.~101115575, 
Neuromorphic Polariton Accelerator -- (PolArt) funding from the Horizon-EIC-2023-Pathfinder Open EU programme under grant agreement No.~101130304. F.T. acknowledges funding from MUR through the PRIN 2022/PNRR project ‘TWo-dimensional materials manIpulation with Surface acousTic wavEs Resonators’ -- TWISTER (grant P2022R7FHC) and the MAECI project ‘DEFEQT - DEterministically FabricatEd Quantum emitters in Two dimensional materials’. Views and opinions expressed are however those of the author(s) only and do not necessarily reflect those of the European Union or European Innovation Council and SMEs Executive Agency (EISMEA). Neither the European Union nor the granting authority can be held responsible for them. 
N.V. and A.G. acknowlegde the financial support from the NRNU MEPhI Priority 2030 Program and the Foundation for the Advancement of Theoretical Physics and Mathematics ``BASIS'', under the Grant No.~22--1--5--30--1 (A.G.).

\bibliographystyle{ieeetr}
\bibliography{biblio} 
\end{document}


\baselineskip24pt
\maketitle

\blfootnote{$^{\dagger}$ These authors contributed equally: L.~Polimeno, F.~Todisco\\
$^{\ast}$ e-mail: laura.polimeno@nanotec.cnr.it, francesco.todisco@nanotec.cnr.it}

\section{Holes pattern}
Circular holes or stripes pattern was etched on the surface of commercial silicon dioxide (300 nm thick) on silicon wafers. The substrates were cleaned by sonication in acetone and 2-propanhol, followed by oxygen plasma (200 W, 100 sccm, 5 minutes). The sample was then spin coated with 300 nm thick polymethyl methacrylate (950K PMMA A4, MicroChem) and soft baked at 180 °C for 15 minutes. Electron beam litography was then carried out using a 250 $\mu C$ beam dose and development in a 2-propanhol water 3:1 mixture at room temperature (30 sec). The holes pattern was finally etched in silicon dioxide by wet etching in a 6:1 buffered oxide etchant for 180 sec (Sigma Aldrich). The PMMA mask was finally removed by immersion in hot acetone for 10 minutes.

\begin{figure}[h!]
\centering \includegraphics[scale=0.25]{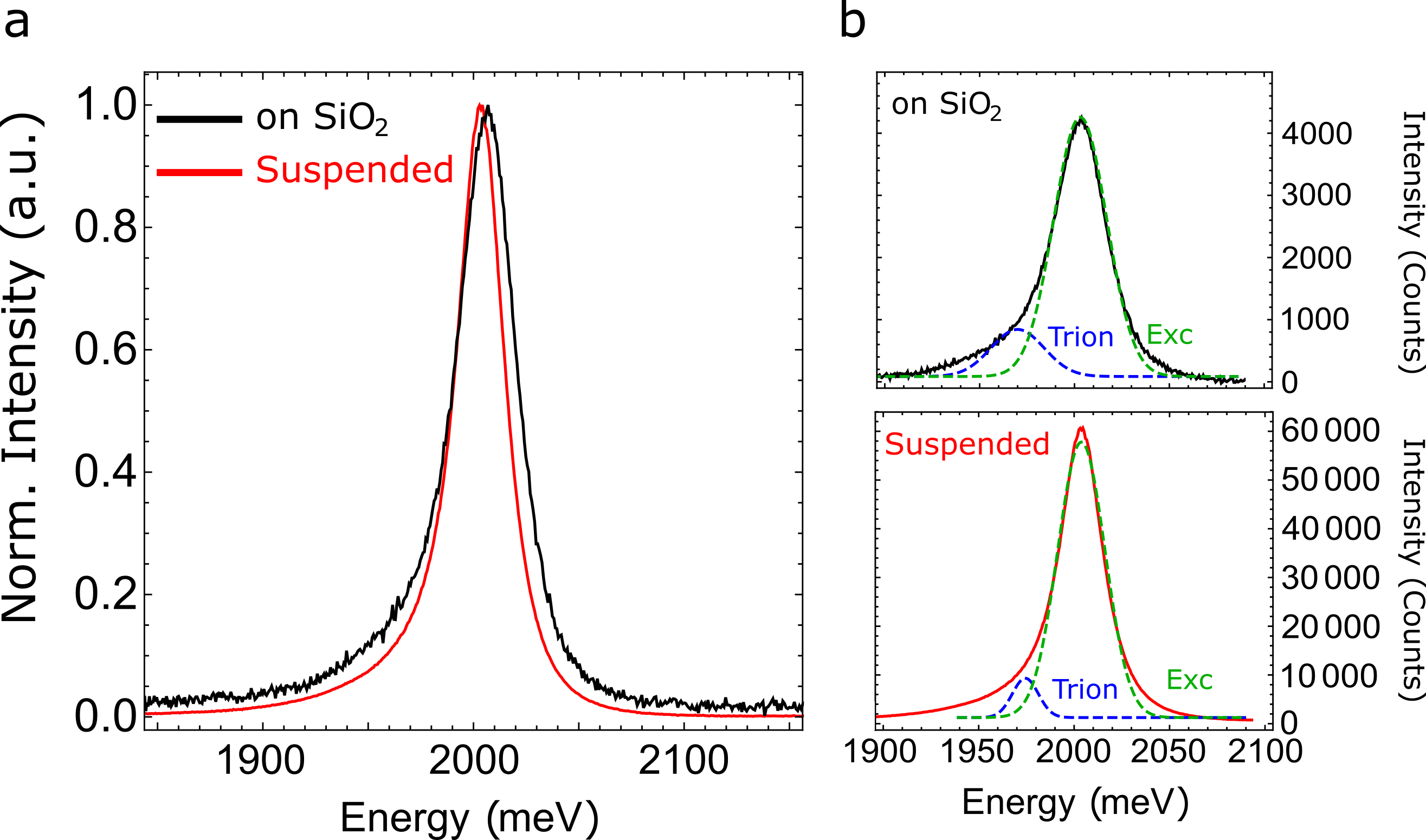}
\caption{a) Comparison of the normalized PL spectra between the suspended region (red continuous line) and the SiO$_2$/Si substrate (black continuous line) for analyzing the PL lineshape. b)  Fit functions of the experimental data for exciton (green dashed line) and trion (blue dashed line) in both cases. }
\label{} 
\end{figure}

\section{Sample fabrication}

\begin{figure}[h!]
\centering \includegraphics[scale=0.65]{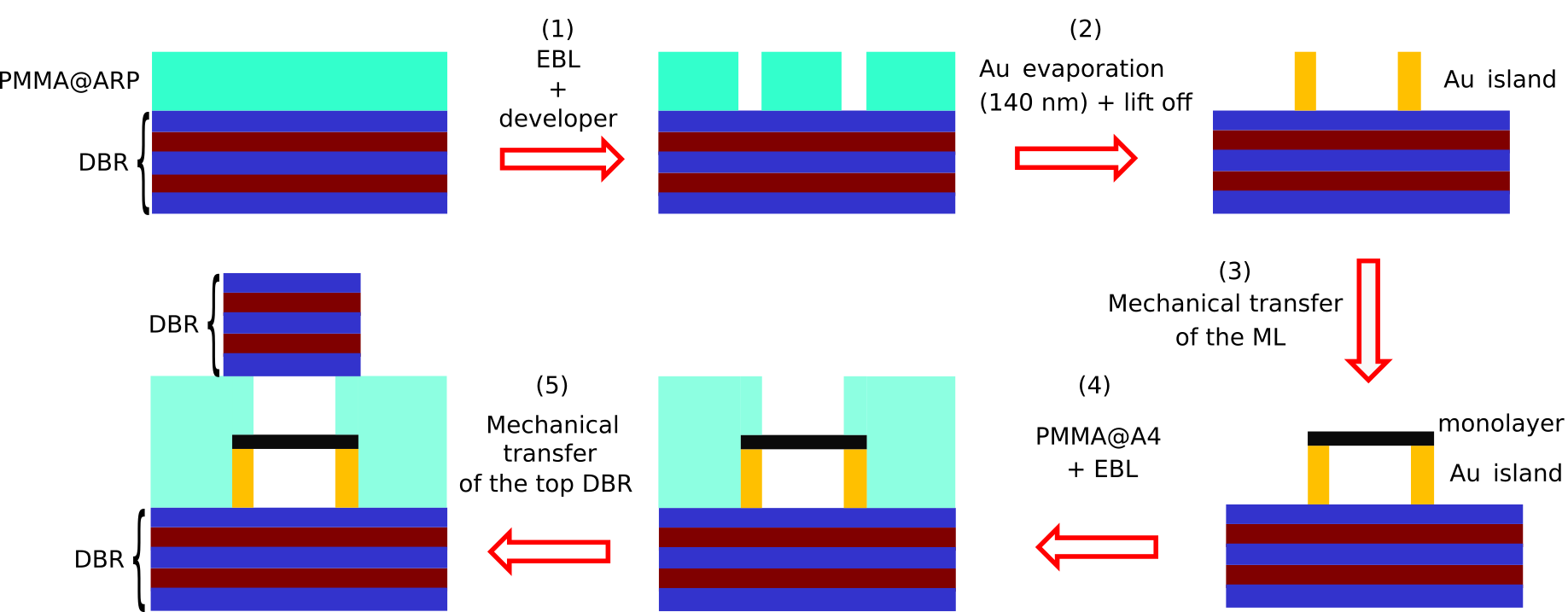}
\caption{Step by step fabrication process of WS$_2$ suspended monolayer embedded in a planar microcavity.}
\label{} 
\end{figure}
Fig. S1 shows, step by step, the fabrication process developed for the realization of the suspended monolayer-based planar microcavity. (1) A 500 nm-thick PMMA ARP671.04 was directly written via Electron Beam Lithography (EBL), imprinting the final design: a 30 $\mu$m x 30 $\mu$m square with two strips with dimensions of 0.5 $\mu$m x 10 $\mu$m and 1 $\mu$m x 10 $\mu$m. (2) A 140 nm-thick gold film is evaporated on the substrate and the lift off process are made in acetone for 2 hours. (3) The WS$_2$ monolayer is mechanically transferred on the gold island, suspending the monolayer region below the two strips. The sample is annealed at 200$^\circ$ in vacuum for 3 hours, for improving the adhesion between the gold and the monolayer (Fig. S2).
\begin{figure}[h!]
\centering \includegraphics[scale=0.6]{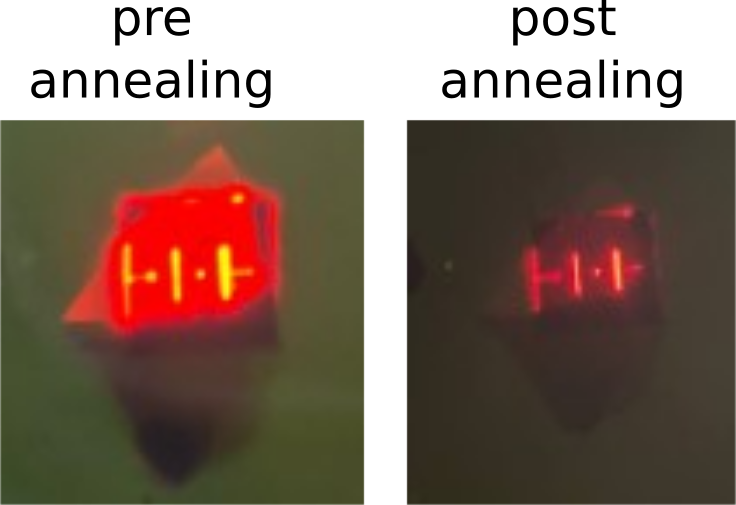}
\caption{Comparison between the monolayer emission after the transfer and after the annealing.}
\label{} 
\end{figure}
(4) For the top part of the cavity, 140 nm-thick PMMA A4 was spun on the top of the monolayer and the EBL writing was aligned with the bottom strips. To optimize the overleap of the suspended area, the top strips were enlarged with respect the first one. The top strips were exposed with a 30kV electron beam, using a positive tone process at low doses; the sample was finally developed with a low-stress/low-dose developer, followed by a surfactant-enhanced deionized water rinsing and $N_2$ drying. (5) The optical microcavity was closed with the transferred of the DBR, via all-dry transfer. 

\section{Dispersion on contacted region}

\begin{figure}[h!]
\centering \includegraphics[scale=0.43]{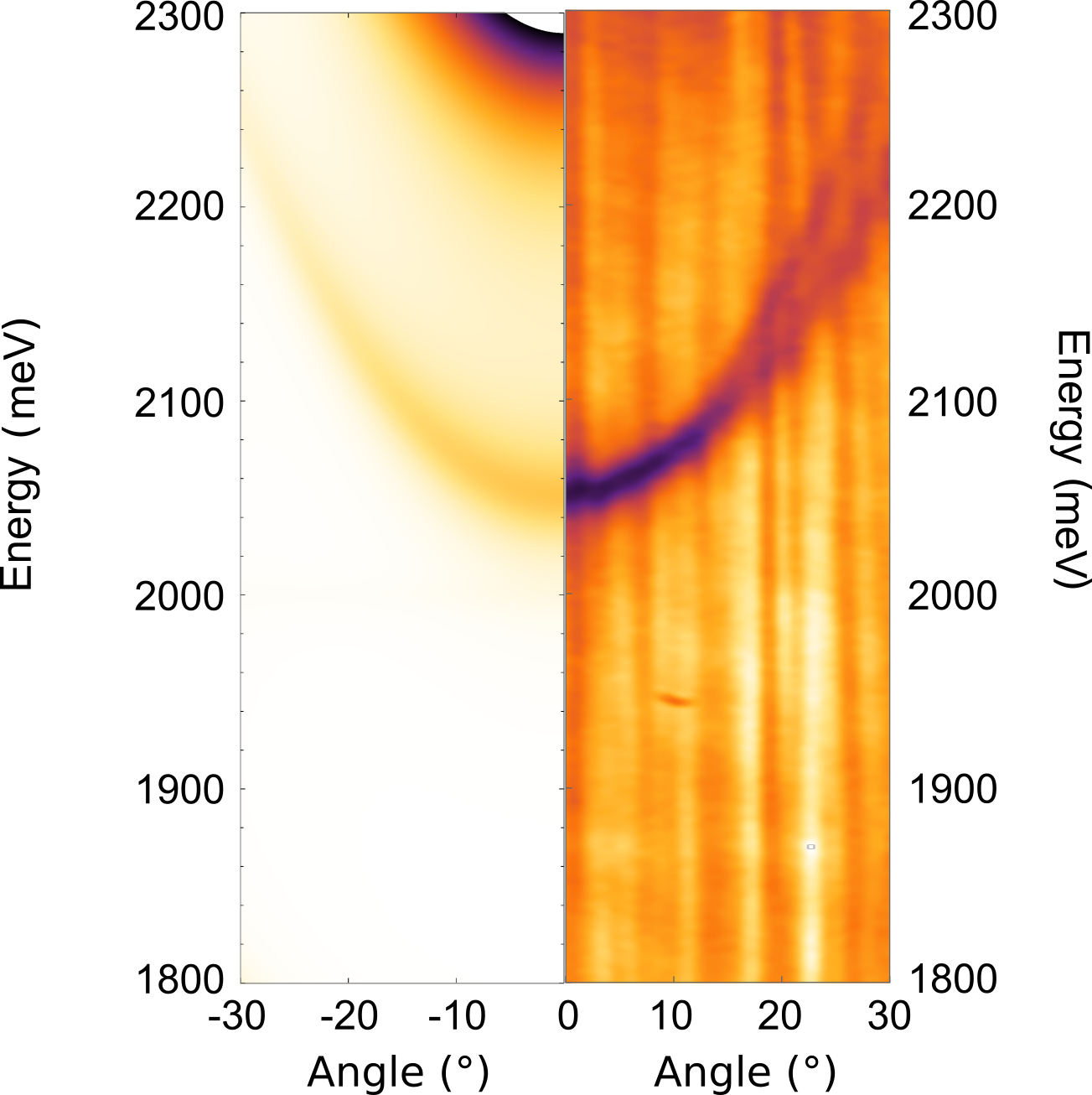}
\caption{d) Comparison between the theoretical simulation and the experimental reflectivity in Fourier space of the contacted region.}
\label{} 
\end{figure}
The reflectivity dispersion in the contacted region of the monolayer indicates the presence of the uncoupled optical mode, as theoretically expected.
In this sample region, the monolayer is in direct contact with the gold island, mostly quenching the exciton emission and removing the air spacer at the bottom of the cavity.

\section{Monolitic microcavity}
To directly compare the spin-dependent interactions of the suspended microcavity with the monolithic one, we measure the blueshift of the lower polariton mode of a typical microcavity at RT as a function of the incident power, by varying the polarization of the excitation (Fig. S6).
\begin{figure}[h!]
\centering \includegraphics[scale=0.43]{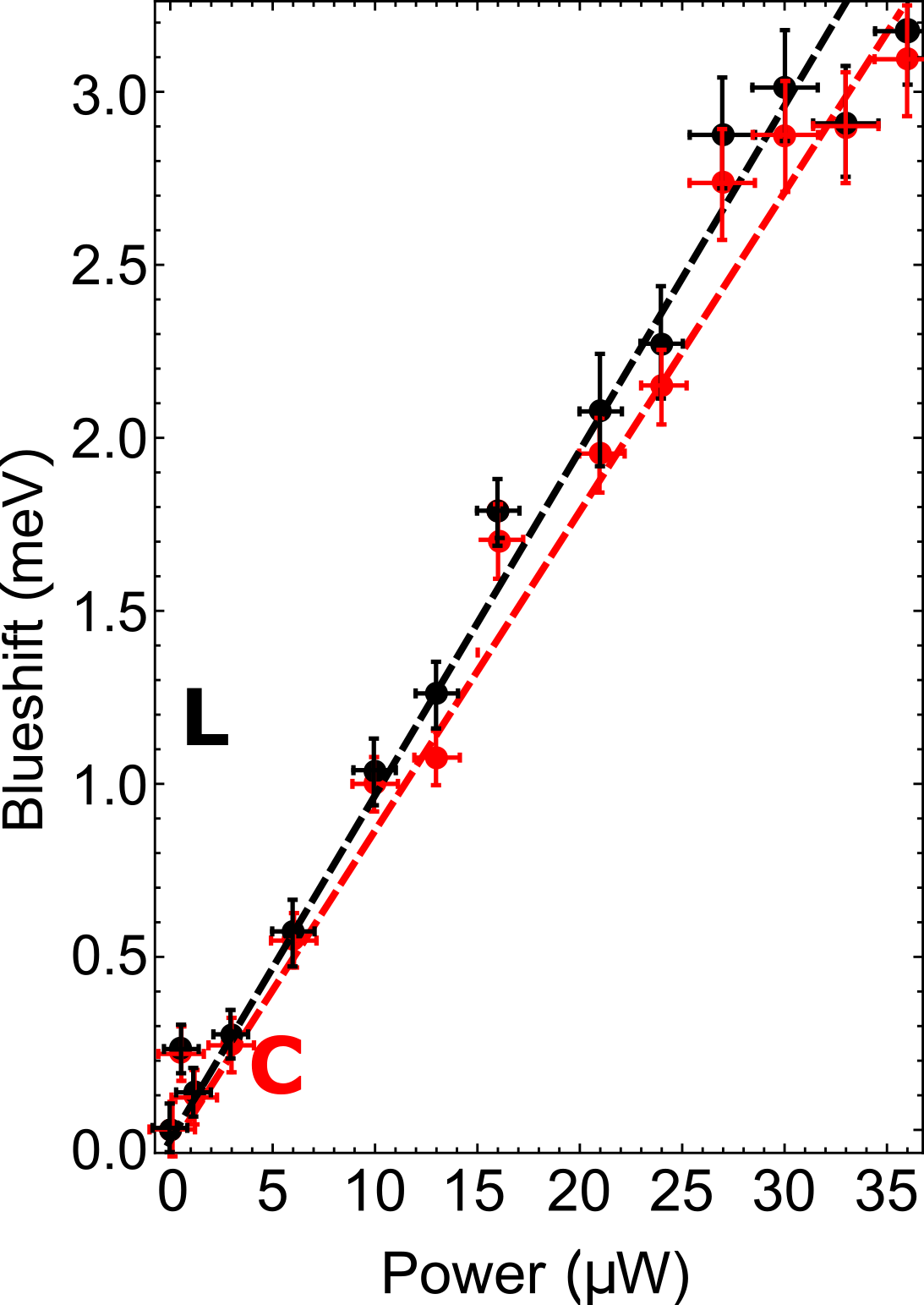}
\caption{Energy blueshift of the lower polariton as a function of the polariton density inside the cavity, in the case of linearly- (L, black dots) and circularly- (C, red dots) polarized 
excitation for a monolitic microcavity.}
\label{} 
\end{figure}
For this set of data, the polariton-polariton interaction strength is g$_{pol}$ $\sim$ 0.005 $\mu$eV $\mu$m$^2$ while the excitonic one is $g_{exc}$ $\sim$ 0.2 $\mu$eV $\mu$m$^2$.

\section{Calculation of interaction constants at finite $T$}

The standard approach to theoretically estimate polariton nonlinearities~\cite{tassone,glazov}, yields the exciton interaction constant 
\begin{equation}\label{gex(0)}
g_{exc}^0 = 2 \int \! \frac{d{\bf p}}{(2\pi)^2}  \int \! \frac{d{\bf q}}{(2\pi)^2} V({\bf q})\chi({\bf p})\chi^{*}({\bf p})\chi^{*}({\bf p} -{\bf q})\bigl[\chi({\bf p} )-\chi({\bf p -q} )\bigr]
\end{equation}
and the saturation constant 
\begin{equation}\label{gsat(0)}
g_{sat}^0 = \frac{\hbar \Omega_{\rm R}}{2\chi({\bf r}=0)} \int \! \frac{d{\bf p}}{(2\pi)^2}\, \chi{(\bf p)}\chi^{*}{(\bf p)}\chi{(\bf p)},
\end{equation}
where $V({\bf q})$ is the Fourier image of electron-hole interaction (the Rytova-Keldysh potential), $\hbar\Omega_{\rm R}$ is the Rabi splitting, $\chi({\bf r})$ and $\chi({\bf p})$ are the $1s$--exciton wavefunction in position and momentum bases, respectively. The expressions in Eqs.~(\ref{gex(0)}) and~(\ref{gsat(0)}) account for electrostatic (direct and exchange) interaction and for the phase-space filling for the excitonic species of the same spin, respectively, and are valid for the case of zero temperature and the so-called rigid excitons which are strictly in $1s$ state. 
While the latter assumption is fulfilled in TMD materials due to large exciton binding energies, the experimental conditions of our experiment are far from $T=0$. 

In this section, we derive the temperature dependencies of $g_{\rm ex}$ and $g_{\rm sat}$ using the previously developed approach to bosonization in polariton systems~\cite{grudinina2024}. 
Our starting point is the Bethe-Salpeter equation (BSE)  
\begin{multline}\label{BSE}
    \left(E_{\rm g} +\frac{\hbar^2 {\bf k}^2}{2 M} +\frac{\hbar^2 {\bf p}^2 }{2m} - \mu - i \Omega\right)\Delta({\bf p},{\bf k}, \Omega) \\-\! \left[n_{v}\!\left(\!{\bf p} - \!\frac{m}{m_h} {\bf k}\!\right)\! - n_{c}\!\left(\!{\bf p}\! + \!\frac{m}{m_e} {\bf k}\!\right)\!\right]\int \!\!\frac{d{\bf q}}{(2\pi)^2}V({\bf p} -{\bf q}) \Delta( {\bf q},{\bf k},\Omega) =0
\end{multline}
where $\Delta({\bf q},{\bf k},\Omega)$ is the exciton field dependent on the exciton c.m. momentum ${\bf k}$ and the relative momentum ${\bf p}$ of the electron-hole motion, $\Omega$ is the bosonic Matsubara frequency, $M$ is the exciton mass, $m$ is the reduced mass of an electron and a hole with respective masses $m_e$ and $m_h$, $E_{\rm g}$ and $\mu$ are the exciton gap and the chemical potential, and $n_{c(v)}$ denote the Fermi distributions of electrons (holes) in the conductance (valence) bands at a given temperature $T$. 

For simplicity, to derive the nonlinear terms in the effective exciton-photon action 
with respect to the BSE, in the Fermi distributions we neglect the c.m. momentum compared to the momentum of internal motion. Then the exciton interaction term reduces to
\begin{multline}
    \mathcal{S}^{(4)}_{\rm ex}\!= \!\frac{T}{2}\!\!\sum_{\substack{{\bf q}, {\bf p},\\{\bf k}_1, {\bf k}_2,{\bf k}_3}} \!\sum_{\Omega_{1}, \Omega_{2}, \Omega_3} 2 V({\bf q}) \frac{\Delta({\bf p}, {\bf k}_1, \Omega_1)  \bar\Delta({\bf p} + \tfrac{{\bf k}_1 - {\bf k}_2}{2}, {\bf k}_2, \Omega_2)}{[n_{v}({\bf p}) - n_{c}({\bf p})]^2} \\
    \times
    \bar\Delta({\bf p}+ \tfrac{{\bf k}_3 - {\bf k}_1}{2}, {\bf k}_3, \Omega_3)\Delta({\bf p} - {\bf q } + \tfrac{{\bf k}_{3} - {\bf k}_2}{2}, {\bf k}_2 + {\bf k}_3-{\bf k}_1, \Omega_{2} + \Omega_3 - \Omega_1) 
    \\[5pt] 
    -\!\frac{T}{2}\!\!\sum_{\substack{{\bf q}, {\bf p},\\{\bf k}_1, {\bf k}_2,{\bf k}_3}} \!\sum_{\Omega_{1}, \Omega_{2}, \Omega_3} 2 V({\bf q}) \frac{\Delta({\bf p}, {\bf k}_1, \Omega_1)  \bar\Delta({\bf p} + \tfrac{{\bf k}_1 - {\bf k}_2}{2}, {\bf k}_2, \Omega_2)}{[n_{v}({\bf p}) - n_{c}({\bf p})][n_{v}({\bf p -q}) - n_{c}({\bf p-q})]}  \\\times
    \Delta({\bf p}-{\bf q} -{\bf k}_2 +\tfrac{{\bf k}_3 + {\bf k}_1}{2}, {\bf k}_3, \Omega_3)\bar \Delta({\bf p} - {\bf q } + \tfrac{{\bf k}_{3} - {\bf k}_2}{2}, {\bf k}_2 + {\bf k}_3-{\bf k}_1, \Omega_{2} + \Omega_3 - \Omega_1),
\end{multline}
while the saturation nonlinearity reads:
\begin{multline}
     \mathcal{S}^{(4)}_{\rm sat}\!= \!\frac{T}{2}\!\!\sum_{\substack{{\bf q}, {\bf p},\\{\bf k}_1, {\bf k}_2,{\bf k}_3}} \!\sum_{\Omega_{1}, \Omega_{2}, \Omega_3} 
 \frac{\hbar\Omega_R}{2\chi({\bf r}=0)} \left[  \frac{\Delta({\bf p}, {\bf k}_1, \Omega_1)  \bar\Delta({\bf p} + \tfrac{{\bf k}_1 - {\bf k}_2}{2}, {\bf k}_2, \Omega_2)}{[n_{v}({\bf p}) - n_{c}({\bf p})]^2} \right.\\
 \Biggl.\times \bar\Delta({\bf p}+ \tfrac{{\bf k}_3 - {\bf k}_1}{2}, {\bf k}_3, \Omega_3)
    \Psi_{\rm ph}({\bf k}_2 + {\bf k}_3 -{\bf k}_1,\Omega_2 + \Omega_3- \Omega_1)  + {\rm c.c.}\Biggr]
\end{multline}
with $\Psi_{\rm ph}$ denoting the photon field.

Under the assumption $|{\bf k}_{i}|\ll |{\bf p}|, |{\bf q}|$ the variables in the BSE (\ref{BSE}) can be separated as $\Delta( {\bf p},{\bf k},\Omega) =\chi({\bf p})C({\bf k}, \Omega)$, where $\chi({\bf p})$ is the $1s$--exciton wavefunction modified by temperature: 
\begin{equation}
    \left(\frac{{\bf p}^2}{2m} - E_{\rm 1s}\right)\chi({\bf p}) = \left[n_{v}\!\left({\bf p} \right) - n_{c}\!\left({\bf p}\right)\right]\int \frac{d{\bf q}}{(2\pi)^2}V({\bf p} -{\bf q})\chi({\bf q}).
\end{equation}
In this case the expressions for the interaction and saturation constants are as follows:
\begin{align}
g_{exc} & \! = \!\!\! \int \!\! \frac{ d{\bf p}}{(2\pi)^2}  \!\!\int \!\!\frac{ d{\bf q}}{(2\pi)^2} V({\bf q})|\chi({\bf p})|^2 \frac{\chi^{*}({\bf p\!-\!q})}{n_{v}({\bf p}) \!-\! n_{c}({\bf p})} \!\! \left( \!\frac{\chi({\bf p})}{n_{v}({\bf p}) \!-\! n_{c}({\bf p})} \!-\! \frac{\chi({\bf p \!-\! q})}{n_{v}({\bf p \!-\! q}) \!-\! n_{c}({\bf p \!-\! q})}\!\right)\label{gex} \\
g_{sat}& =\frac{\hbar \Omega_R}{2 \chi(r=0)} \int\frac{d{\bf p}}{(2\pi)^2} |\chi({\bf p})|^2 \frac{\chi({\bf p})}{[n_{v}({\bf p}) - n_{c}({\bf p})]^2}\label{gsat}
\end{align}

\begin{figure}[t]
\centering 
\includegraphics[width=\linewidth]{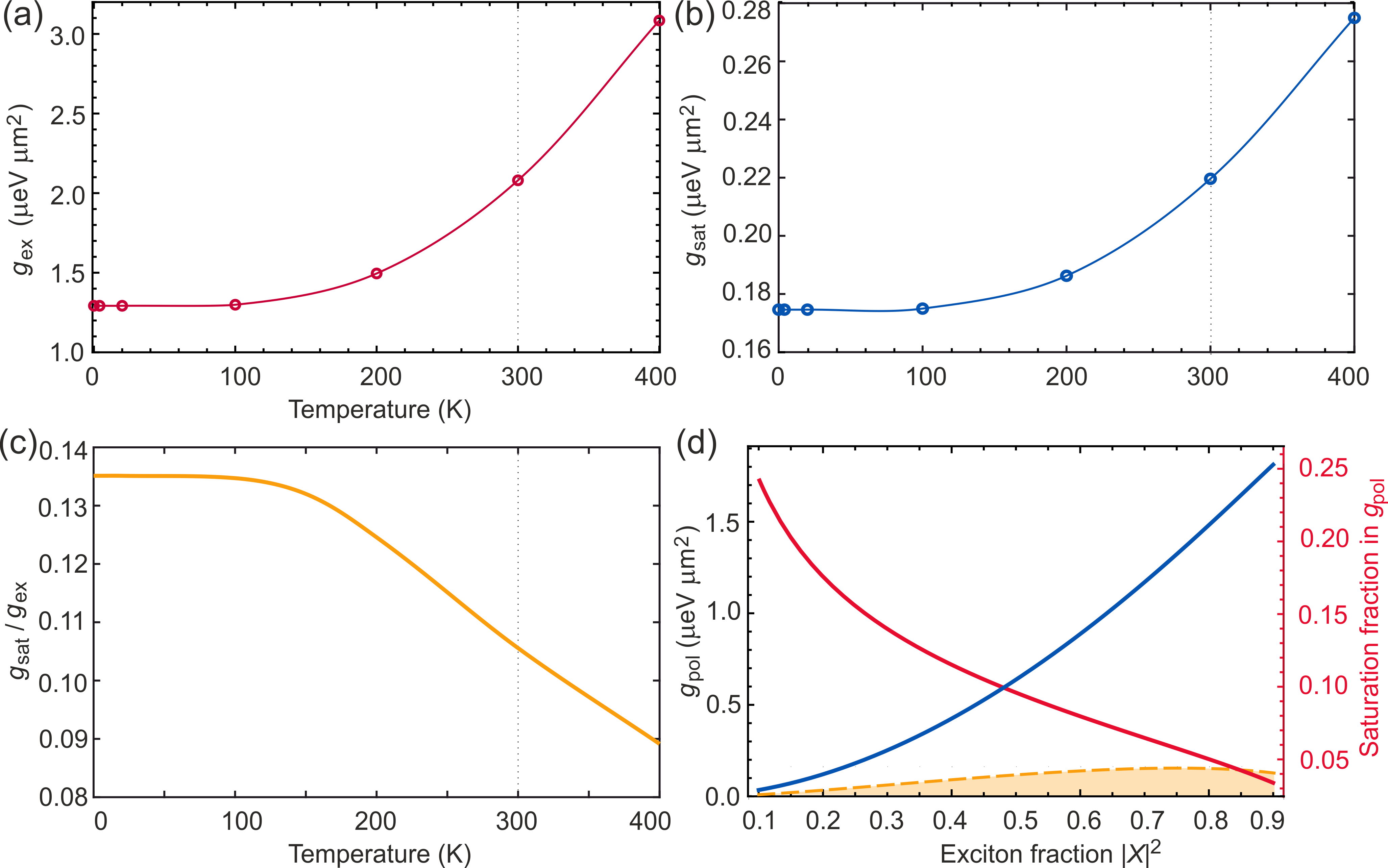}
\caption{(a) Exciton interaction $g_{exc}$ and (b) saturation $g_{sat}$ constants vs. temperature according to Eqs.~(\ref{gex})--(\ref{gsat}). Parameters: $m_e = 0.27 m_0$, $m_h = 0.36 m_0$ \cite{Kormanyos_2015}, $E_{\rm g}^0 =2.63$~eV, $E_{\rm g}(T) =E_{\rm g}^0 - 0.53 T^2/(T + 118.9)$ \cite{Godiksen}, $E_{\rm b} = 518$~meV, chemical potential $\mu=\mu_e+\mu_h$, $\mu_e = \mu_h=1.2$~eV. 
(c) The ratio $g_{sat}/g_{exc}$ vs. temperature, showing the increasing role of the interaction contribution to the nonlinearity with the growth of $T$. (d) The polariton nonlinearity $g_{pol}$ at RT calculated according to Eq.~(\ref{gpol}) vs. the excitonic fraction in the lower polariton (blue line) and the contribution to it from exciton saturation (yellow shading). Right axis, red line: fraction of the blueshift coming from saturation.}
\label{int_const} 
\end{figure}

\noindent The renormalization of the interaction constants due to finite temperature is introduced in the expressions above via the Fermi distribution functions for electron and holes. Finite temperatures lead as well to smearing of the excitonic wavefunction according to Eq.~(\ref{BSE}) and, as a result, to the increase of the overlap integral in Eq.(\ref{gex})--(\ref{gsat}), which results in the effective growth of the interaction constants with $T$.

In Fig.~\ref{int_const} we plot the temperature dependence of the two exciton-polariton nonlinearities for the case when the chemical potentials of electrons and holes lie approximately in the middle of the bandgap. One sees that while for cryogenic temperatures (below 100~K) the difference from the zero-temperature values (\ref{gex(0)})--(\ref{gsat(0)}) is negligible, with the further increase of temperature towards ambient conditions, as soon as $k_BT$ becomes comparable with the exciton binding energy $E_{\rm b}$, the expected values for $g_{exc}$ and $g_{sat}$ start to grow. We note that at the same time that for all temperatures in consideration, the saturation constant stays one order of magnitude lower than $g_{\rm ex}$, and the ratio between them starts to linearly decrease at approximately the same temperatures ($\sim100$~K, see Fig.~\ref{int_const}c).

The theoretical lower-polariton blueshift reads (see~\cite{grudinina2024} for details)
\begin{equation}\label{blueshift}
    \Delta E_{pol} = g_{exc} n_{exc} |X|^2 + 2 g_{sat} n_{exc}|X|\sqrt{1-|X|^2},
\end{equation}
where $n_{exc}$ is the exciton density and $X$ is the exciton Hopfiled coefficient (in our case, taken at zero wavevector) which is defined from the experimental Rabi splitting and detuning $\delta$ as
$|X|^2 = (1-\delta/\sqrt{\hbar\Omega_R^2 + \delta^2})/2$. Comparing with the experimentally measured $\Delta E_{pol} = g_{pol}\cdot n_{pol}$ and given $n_{exc} = |X|^2n_{pol}$ ($n_{pol}$ is the experimentally-assessed polariton density), we conclude
\begin{equation}\label{gpol}
    g_{pol} = g_{exc}|X|^4 + 2g_{sat}|X|^3\sqrt{1-|X|^2}.
\end{equation}
From Eq.~(\ref{gpol}), in the situations when one can neglect the saturation-related term, one arrives at the common approximation $\Delta E_{pol}\approx   g_{exc}|X|^4n_{pol}$ used to obtain the experimental lower bound  $g_{exc}^{\rm RT}=1.65~\mu$eV~$\mu$m$^2$ for the case of circularly-polarized excitation, which is in good correspondence with our theoretically estimated value $g_{exc}^{\rm RT}\approx2.08~\mu$eV~$\mu$m$^2$. 

Finally, in Fig.~\ref{int_const}d we report the behaviour of the effective polariton nonlinearity $g_{pol}$ and its contribution from exciton saturation depending on the exciton fraction $|X|^2$. As one can see, with the growth of $|X|^2$ the role of saturation quickly diminishes, thus one can expect that going towards zero or positive detunings (resulting in $|X|^2>0.5$) will increase the polariton nonlinearity by more than an order of magnitude.


As a side-note, it needs to be commented that the commonly used for zero-temperature WS$_2$ theoretical approximation~\cite{shahnazaryan2017exciton,barachati2018interacting} $g_{exc}^0 = 2.07E_{\rm b}\lambda^{2}_{x}$ in terms of the exciton binding energy and the $1s$-exciton Bohr radius $\lambda_{x}$ proves to be misleading. The prefactor $2.07$ obtained using the hydrogenic ansatz for the exciton wavefunction $\sim\exp\{-r/\lambda_x\}$ is erroneous as it was obtained using a wrong value of the screening length of the material $\rho_0=7.5$~nm in their variational procedure. Given the correct value $\rho_0 = 2\pi\alpha=3.76$~nm, where $\alpha=0.6$ is polarizability of WS$_2$ (see also~\cite{berkelbach,crooker}), 
and the Bohr radius $\lambda_x = 1.7$~nm, with the hydrogenic ansatz substituted to Eq.~(\ref{gex(0)}) one gets $g_{exc}^0 = 3.2~\mu$eV~$\mu$m$^2$ (cf. the value $1.9~\mu$eV~$\mu$m$^2$ reported in~\cite{shahnazaryan2017exciton,barachati2018interacting}). Furthermore, substitution of the accurate (numerically simulated) shape of the exciton wavefunction in Eq.~(\ref{gex(0)}) yields $g_{exc}^0=1.2923~\mu$eV~$\mu$m$^2$ which underlines as well that the hydrogenic ansatz is not to be used (it provides a good estimate for the binding energy but not for the wavefunction shape).

\bibliographystyle{ieeetr}
\bibliography{biblio}